\input{aipcheck}

\documentclass[
    ,final            
  ]
  {aipproc}

\layoutstyle{6x9}

\usepackage{bm}
\newcommand{\br}{\bm{r}}

\begin{document}

\title{Microscopic and self-consistent description for neutron halo in deformed nuclei}


\author{Lulu Li}{
  address={Institute of Applied Physics and Computational Mathematics, Beijing 100094, China}
}

\author{Jie Meng}{
  address={State Key Laboratory of Nuclear Physics and Technology, School of Physics, 
              Peking University, Beijing 100871, China}
  ,altaddress={School of Physics and Nuclear Energy Engineering, 
              Beihang University, Beijing 100191, China}
  ,altaddress={Department of Physics, University of Stellenbosch, Stellenbosch, South Africa}
}

\author{P. Ring}{
  address={Physikdepartment, Technische Universit\"at M\"unchen,
              85748 Garching, Germany}
  ,altaddress={State Key Laboratory of Nuclear Physics and Technology, School of Physics, 
              Peking University, Beijing 100871, China}
}

\author{En-Guang Zhao}{
  address={State Key Laboratory of Theoretical Physics,
              Institute of Theoretical Physics, Chinese Academy of Sciences,
              Beijing 100190, China}
  ,altaddress={Center of Theoretical Nuclear Physics, National Laboratory
              of Heavy Ion Accelerator, Lanzhou 730000, China} 
}

\author{Shan-Gui Zhou}{
  address={State Key Laboratory of Theoretical Physics,
              Institute of Theoretical Physics, Chinese Academy of Sciences,
              Beijing 100190, China}
  ,altaddress={Center of Theoretical Nuclear Physics, National Laboratory
              of Heavy Ion Accelerator, Lanzhou 730000, China} 
}

\begin{abstract}
A deformed relativistic Hartree-Bogoliubov theory in continuum 
has been developed for the study of neutron halos in deformed nuclei
and the halo phenomenon in deformed weakly bound nuclei is investigated. 
Magnesium and neon isotopes are studied and some results are 
presented for the deformed neutron-rich and weakly bound nuclei $^{44}$Mg and $^{36}$Ne. 
The core of the former nucleus is prolate, but the halo has a slightly oblate shape. 
This indicates a decoupling of the halo orbitals from the deformation of the core. 
The generic conditions for the existence of halos in deformed nuclei and 
for the occurrence of this decoupling effect are discussed.
\end{abstract}

\maketitle


\section{\label{sec:intro}Introduction}

The halo phenomenon is one of the most interesting topics in modern nuclear physics. 
Since most open shell nuclei are deformed, the interplay between the deformation 
and the weak binding feature raises interesting questions. 
In order to give an adequate description of possible halo in a deformed nucleus, 
a model should be used which includes in a self-consistent way the continuum, 
large spatial distributions, deformation effects, and couplings among all these features. 
For this purpose, a deformed relativistic Hartree-Bogoliubov theory in continuum 
has been developed and the halo phenomenon 
in deformed weakly bound nuclei is investigated. 

\section{\label{sec:formalism} Theoretical framework}

\subsection{Continuum contribution and relativistic Hartree-Bogoliubov theory}

By solving the non-relativistic
Hartree-Fock-Bogoliubov (HFB)~\cite{Bulgac1980_nucl-th9907088, Dobaczewski1984_NPA422-103,
Dobaczewski1996_PRC53-2809} or the relativistic Hartree Bogoliubov
(RHB)~\cite{Meng1996_PRL77-3963, Poschl1997_PRL79-3841, Lalazissis1998_PLB418-7,
Meng1998_NPA635-3, Meng2006_PPNP57-470} equations in coordinate ($r$) space,
the mean field effects of the coupling to the continuum can be fully taken into account.
With the relativistic continuum Hartree-Bogoliubov (RCHB) theory~\cite{Meng1998_NPA635-3, 
Meng2006_PPNP57-470}, 
properties of the halo nucleus $^{11}$Li has been reproduced quite well~\cite{Meng1996_PRL77-3963}
and the prediction of giant halos in light and medium-heavy nuclei was made~\cite{Meng1998_PRL80-460,
Meng2002_PRC65-041302R, Zhang2003_SciChinaG46-632}.
The RCHB theory has been generalized to treat the odd particle system~\cite{Meng1998_PLB419-1}
and combined with the Glauber model, 
the charge-changing cross sections for C, N, O and F isotopes on a carbon target
have been reproduced well~\cite{Meng2002_PLB532-209}.

The Dirac Hartree Bogoliubov (RHB) equation for the nucleons reads~\cite{Kucharek1991_ZPA339-23},
\begin{eqnarray}
 \int d^3 \bm{r}'
 \left(
  \begin{array}{cc}
   h_D
   - \lambda &
   \Delta
   \\
  -\Delta^*
   & -h_D
   + \lambda \\
  \end{array}
 \right)
 \left(
  { U_{k}
  \atop V_{k}
   }
 \right)
 & = &
 E_{k}
  \left(
   { U_{k}
   \atop V_{k}
    }
  \right)
 ,
 \label{eq:RHB0}
\end{eqnarray}
where $E_{k}$ is the quasiparticle energy, $\lambda$ is the chemical potential,
and $h_D$ is the Dirac Hamiltonian,
\begin{equation}
 h_D(\bm{r}, \bm{r}') =
  \bm{\alpha} \cdot \bm{p} + V(\bm{r}) + \beta (M + S(\bm{r})).
\label{eq:Dirac0}
\end{equation}
with scalar and vector potentials
\begin{eqnarray}
S(\bm{r}) & = & g_\sigma \sigma(\bm{r}), \label{eq:vaspot}\\
V(\bm{r}) & = & g_\omega \omega^0(\bm{r}) +g_\rho \tau_3 \rho^0(\bm{r})
                    +e \displaystyle\frac{1-\tau_3}{2} A^0(\bm{r}) .
\label{eq:vavpot}
\end{eqnarray}
The equations of motion for the mesons and the photon
\begin{eqnarray}
 \left\{
   \begin{array}{rcl}
    \left( -\Delta + \partial_\sigma U(\sigma) \right )\sigma(\bm{r})
      & = & -g_\sigma \rho_s(\bm{r}) , \\
    \left( -\Delta + m_\omega^2 \right )             \omega^0(\bm{r})
      & = &  g_\omega \rho_v(\bm{r}) , \\
    \left( -\Delta + m_\rho^2 \right)                  \rho^0(\bm{r})
      & = &  g_\rho   \rho_3(\bm{r}) , \\
    -\Delta                                               A^0(\bm{r})
      & = &  e        \rho_p(\bm{r}) ,
   \end{array}
 \right.
 \label{eq:mesonmotion}
\end{eqnarray}
have as sources the various densities
\begin{eqnarray}
 \left\{
  \begin{array}{rcl}
   \rho_s(\bm{r})
   & = &
    \sum\limits_{k>0} V_{k}^\dagger(\bm{r})\gamma_0 V_{k}(\bm{r}) ,\\
   \rho_v(\bm{r})
   & = &
    \sum\limits_{k>0} V_{k}^\dagger(\bm{r}) V_{k}(\bm{r}) ,\\
   \rho_3(\bm{r})
   & = &
    \sum\limits_{k>0} V_{k}^\dagger(\bm{r}) \tau_3 V_{k}(\bm{r}) ,\\
   \rho_c(\bm{r})
   & = &
    \sum\limits_{k>0} V_{k}^\dagger(\bm{r})
                 \displaystyle\frac{1-\tau_3}{2}V_{k}(\bm{r}) ,
  \end{array}
 \right.
 \label{eq:mesonsource}
\end{eqnarray}
where, according to the no-sea approximation, the sum over $k>0$ runs
over the quasi-particle states corresponding to single particle
energies in and above the Fermi sea.

In the particle-particle (pp) channel, we use a density dependent  zero range force,
\begin{equation}
 V^\mathrm{pp}(\br_1,\br_2) =  V_0 \frac{1}{2}(1-P^\sigma)\delta( \mathbf{r}_1 - \mathbf{r}_2 )
   \left(1-\frac{\rho(\br_1)}{\rho_\mathrm{sat}}\right).
 \label{eq:pairing_force}
\end{equation}
$\frac12(1-P^\sigma)$ projects onto spin $S=0$ component in the pairing field.
The pairing potential then reads,
\begin{equation}
 \Delta(\br)=V_0(1-\rho(\br)/\rho_{\rm sat})\kappa(\br) ,
\end{equation}
and we need only the local part of the pairing tensor
\begin{equation}
 \kappa(\br)= \sum_{k>0} V_{k}^\dagger(\bm{r})U^{}_{k}(\bm{r}) .
\label{E12}
\end{equation}

\subsection{Large spatial density distribution and the Woods-Saxon basis}

In order to (1) consider properly the asymptotic behavior of nuclear densities
at large $r$ and (2) make the numerical procedure less complicated, 
the Woods-Saxon basis has been proposed in Ref.~\cite{Zhou2003_PRC68-034323}
as a reconciler between the harmonic oscillator basis and the integration
in coordinate space. 
Woods-Saxon wave functions have a
much more realistic asymptotic behavior at large $r$ than
do the harmonic oscillator wave functions.
On one hand, one can still use a large box boundary condition
to discretize the continuum and easily find numerical solutions for a spherical 
Woods-Saxon potential in $r$ space. 
One the other hand, these Woods-Saxon wave functions
can thus be used as a complete basis for spherical or deformed
systems, and one finally comes back to the familiar
matrix diagonalization problem.
It has been shown that the Woods-Saxon basis can satisfactorily
reproduce the large neutron density distribution in weakly bound
nuclei obtained in $r$ space~\cite{Zhou2003_PRC68-034323}.
Recently, for spherical systems, both non relativistic
and relativistic Hartree-Fock-Bogoliubov theories with forces
of finite range have been developed in a Woods-Saxon
basis~\cite{Schunck2008_PRC78-064305,Long2010_PRC81-024308}.

\subsection{Deformed relativistic Hartree Bogoliubov theory in continuum}

The Woods-Saxon basis can be extended to more complicated situations for 
exotic nuclei where both deformation and pairing have to be taken into account.
Over the past years, lots of efforts have been made to develop a deformed relativistic
Hartree theory~\cite{Zhou2006_AIPCP865-90} and a deformed relativistic Hartree Bogoliubov
theory in continuum (the DefRHBC theory)~\cite{Meng2003_NPA722-C366, Zhou2008_ISPUN2007, 
Zhou2010_PRC82-011301R, Zhou2011_JPCS312-092067, Li2012_PRC85-024312}. 

For axially deformed nuclei with the spatial reflection symmetry, we
expand the potentials $S(\bm{r})$ and $V(\bm{r})$ in Eq.~(\ref{eq:Dirac0})
and various densities in terms of the Legendre
polynomials~\cite{Price1987_PRC36-354},
\begin{equation}
 f(\bm{r})   = \sum_\lambda f_\lambda({r}) P_\lambda(\cos\theta),\
 \lambda = 0,2,4,\cdots
 ,
 \label{eq:expansion}
\end{equation}
with an explicit definition of $f_\lambda({r})$.

The quasiparticle wave functions $U_k$ and $V_k$ in Eq.~(\ref{eq:RHB0}) are  
expanded in the Woods-Saxon basis~\cite{Zhou2003_PRC68-034323}:
\begin{eqnarray}
 U_{k} (\br{s} p)
 & = & \displaystyle
 \sum_{n\kappa} u^{(m)}_{k,(n\kappa)}     \varphi_{n\kappa m}(\br{s} p),
 \label{eq:Uexpansion0} \\
 V_{k} (\br{s} p)
 & = & \displaystyle
 \sum_{n\kappa} v^{(m)}_{k,(n\kappa)} \bar\varphi_{n\kappa m}(\br{s} p).
\label{eq:Vexpansion0}
\end{eqnarray}
$\bar\varphi_{n\kappa m}(\br{s} p)$ is the time reversal state of $\varphi_{n\kappa
m}(\br{s} p)$. 
Because of the axial symmetry the $z$-component $m$ of
the angular momentum $j$ is a conserved quantum number and the RHB Hamiltonian can
be decomposed into blocks characterized by $m$ and parity $\pi$. 
For each $m^\pi$-block, solving the RHB equation (\ref{eq:RHB0}) is
equivalent to the diagonalization of the matrix
\begin{equation}
 \left( \begin{array}{cc}
  {\cal A}-\lambda & {\cal B} \\
  {\cal B^\dag} & -{\cal A}^\ast+\lambda \\
 \end{array} \right)
 \left(
  { {\cal U}_k
    \atop
    {\cal V}_k
  }
 \right)
 = E_k
 \left(
  { {\cal U}_k
    \atop
    {\cal V}_k
  }
 \right),
 \label{eq:RHB1}
\end{equation}
where
\begin{equation}
 {\cal U}_k = \left(u^{(m)}_{k,(n\kappa)}\right),\
 {\cal V}_k = \left(v^{(m)}_{k,(n\kappa)}\right),
\end{equation}
and
\begin{eqnarray}
 {\cal A}
 & = &
 \left( h^{(m)}_{D(n\kappa)(n'\kappa')} \right)
 = 
 \left( \langle n\kappa m|h_D|n'\kappa',m\rangle \right) ,
 \\
 {\cal B}
 & = &
 \left( \Delta^{(m)}_{(n\kappa)(n'\kappa)} \right)~
 = 
 \left( \langle n\kappa m |\Delta| \overline{n'\kappa',m} \rangle \right).
\label{eq:pairing_matrix}
\end{eqnarray}
Further details are given in the appendixes of Ref.~\cite{Li2012_PRC85-024312}.

In order to describe the exotic nuclear structure in unstable odd-$A$ 
or odd-odd nuclei, the DefRHBC theory has been extended to incorporate 
the blocking effect due to the odd nucleon(s)~\cite{Li2012_CPL29-042101}
The deformed relativistic Hartree-Bogoliubov theory in continuum with 
the density-dependent meson-nucleon couplings is developed 
recently~\cite{Chen2012_PRC85-067301}.

\section{\label{sec:results}Results and discussions}

\begin{figure}[hbt!]
\includegraphics[width=0.60\columnwidth]{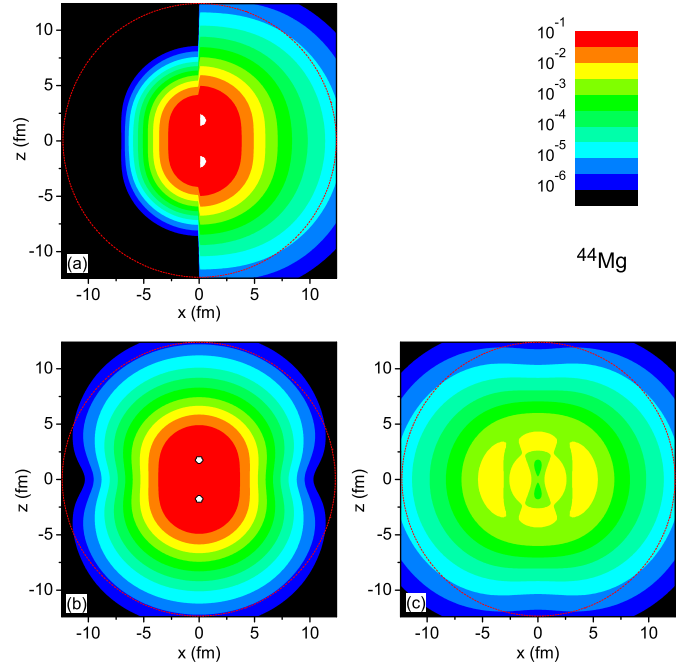}
\caption{
(Color online) Density distributions of $^{44}$Mg with the $z$-axis
as symmetry axis: (a) the proton density (for $x<0$) and the neutron
density (for $x>0$), (b) the density of the neutron core, and (c)
the density of the neutron halo. In each plot, a dotted circle is
drawn for guiding the eye. 
This figure is originally published in Ref.~\cite{Zhou2010_PRC82-011301R}.
}
\label{fig:44Mg}
\end{figure}

We next present some results from the DefRHBC theory 
by taking magnesium and neon isotopes as 
examples and discuss some results for the deformed neutron-rich and 
weakly bound nuclei $^{44}$Mg and $^{36}$Ne~\cite{Zhou2010_PRC82-011301R, 
Zhou2011_JPCS312-092067, Li2012_PRC85-024312}. 

Magnesium isotopes have been studied extensively in 
Refs.~\cite{Zhou2010_PRC82-011301R, Zhou2011_JPCS312-092067, Li2012_PRC85-024312} with the deformed
relativistic Hartree-Bogoliubov theory in continuum and the parameter sets
NL3~\cite{Lalazissis1997_PRC55-540} and PK1~\cite{Long2004_PRC69-034319}. 
For the pp interaction~(\ref{eq:pairing_force}), the following parameters are
used: $\rho_\mathrm{sat} =$ 0.152~fm$^{-3}$ and $V_0 =
380$~MeV$\cdot$fm$^3$, and a cut-off energy
$E^\mathrm{q.p.}_\mathrm{cut} = 60$~MeV is applied in the
quasi-particle space. These parameters were fixed by reproducing the
proton pairing energy of the spherical nucleus $^{20}$Mg obtained
from a spherical relativistic Hartree-Bogoliubov calculation with the
Gogny force D1S. A spherical box of the size $R_\mathrm{max} = 20$ fm
and the mesh size $\Delta r = 0.1$ fm are used for generating the
spherical Dirac Woods-Saxon basis~\cite{Zhou2003_PRC68-034323} which consists of
states with $j< \frac{21}{2} \hbar$. An energy cutoff
$E^+_\mathrm{cut}$ = 100 MeV is applied to truncate the positive
energy states in the Woods-Saxon basis and the number of negative
energy states in the Dirac sea is taken to be the same as that of
positive energy states in each ($\ell,j$)-block.

In our deformed RHB calculations with the parameter set NL3, $^{46}$Mg is
the last nucleus of which the neutron Fermi surface is negative and
the two neutron separation energy is positive~\cite{Zhou2010_PRC82-011301R}.
In the calculations based on the parameter set PK1, $^{42}$Mg is the last
bound nucleus in Mg isotopes~\cite{Li2012_PRC85-024312}.

\begin{figure}
\includegraphics[width=0.6\textwidth]{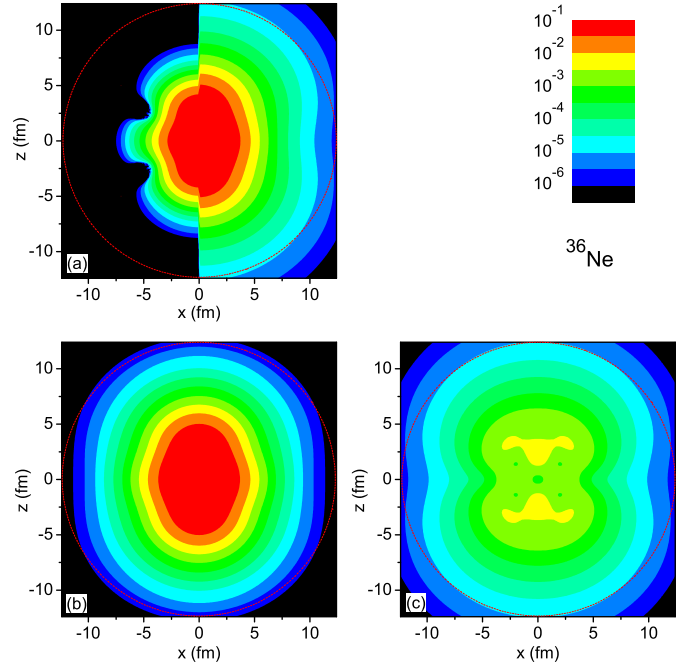}
\caption{\label{fig:36Ne}%
(Color online) Density distributions of $^{36}$Ne. Details are given
in Fig.~\ref{fig:44Mg}.
This figure is originally published in Ref.~\cite{Zhou2010_PRC82-011301R}.
}
\end{figure}

It was found in Ref.~\cite{Zhou2010_PRC82-011301R} that 
the ground state of $^{44}$Mg is well deformed with
quadrupole deformation $\beta_2 = 0.32$
and a very small two neutron separation energy $S_{2n} = 0.44$ MeV.
In the tail part, the neutron density
extends more along the direction perpendicular to the symmetry axis.
The density distribution is decomposed into contributions of the oblate 
``halo'' and of the prolate ``core'' as shown in Fig.~\ref{fig:44Mg}.
The density distribution of this weakly bound nucleus has a very long tail
in the direction perpendicular to the symmetry axis
which indicates the prolate nucleus $^{44}$Mg has an oblate halo and
there is a decoupling between the deformations of the core and the halo.

As discussed in Refs.~\cite{Zhou2010_PRC82-011301R, Zhou2011_JPCS312-092067, Li2012_PRC85-024312},
the shape of the halo originates from the intrinsic structure of the weakly
bound or continuum orbitals. It turns out that in $^{44}$Mg, 
the essential level of the halo has a large contribution from
the prolate $\Lambda = 1$ ($p$ wave) component.
In Ref.~\cite{Zhou2010_PRC82-011301R} an example was also discussed
in which the halo and the core have similar shapes.
In Fig.~\ref{fig:36Ne}a the density distributions of all
protons and all neutrons in the prolate deformed nucleus $^{36}$Ne
are shown ($\beta_2 = 0.52$). It can be seen that
the neutron density not only extends much
farther in space but it also shows a halo structure. The neutron
density is decomposed into the contribution of the core in
Fig.~\ref{fig:36Ne}b and that of the halo in Fig.~\ref{fig:36Ne}c. In
contrary to the nucleus $^{44}$Mg, we observe now a prolate halo,
because the essential level of the halo has a large contribution from
the prolate $\Lambda = 0$ ($p$ wave) component.

The halo feature is connected with relatively large cross sections 
and narrow longitudinal momentum distributions in knockout reactions.
The decoupling between the deformations of the core and the halo
may manifest itself by some new experimental observables, e.g.,
the double-hump shape of longitudinal momentum distribution in 
single-particle removal reactions and new dipole modes, etc.
In particular, a combination of the experimental method proposed in 
Ref.~\cite{Navin1998_PRL81-5089} and the theoretical approach developed 
in Ref.~\cite{Sakharuk1999_PRC61-014609} would be useful in the study of 
longitudinal momentum distribution in single-particle removal reactions 
with deformed halo nuclei as projectiles.
The shape decoupling effects may also has some influence on the 
sub-barrier capture process in heavy ion 
collisions~\cite{Sargsyan2011_2012}.

For odd particle system, the formation and the size of a halo depend
on the interplay among the odd-even effects, continuum and pairing effects,
deformation effects, etc. Some progress on this topic has been made recently~\cite{Li_in-prep}.

\section{Summary}

We present recent progresses of the development of 
a deformed relativistic Hartree-Bogoliubov theory in continuum (DefRHBC) 
and the study of neutron halo in deformed nuclei.
In the very neutron-rich deformed nucleus $^{44}$Mg, pronounced deformed
neutron halo was found. 
The halo is formed by several orbitals close to the threshold. 
These orbitals have large components of low $\ell$-values
and feel therefore only a small centrifugal barrier. 
Although $^{44}$Mg and its cores is prolately deformed, 
the deformation of the halo is slightly oblate. 
This implies a decoupling between the shapes of the core and the halo. 
The mechanism is investigated and 
it was concluded that the existence and the deformation of a possible neutron halo 
depends essentially on the quantum numbers of the main components of the single particle orbits
in the vicinity of the Fermi surface.

\begin{theacknowledgments}
This work has been supported by the Major State Basic
Research Development Program of China (973 Program: ``New and technology at the limits of nuclear
stability''), National Natural Science Foundation of China
(Grants Nos. 10875157, 10975100, 10979066, 11105005, 11175002, 11175252,
11121403, and 11120101005), the Knowledge Innovation Project of Chinese Academy of
Sciences (Grants No. KJCX2-EW-N01 and No. KJCX2-YW-N32), 
and the DFG cluster of excellence \textquotedblleft Origin and Structure of the
Universe\textquotedblright\ (www.universe-cluster.de). 
The results described in this work were obtained
on the ScGrid of Supercomputing Center, Computer
Network Information Center of Chinese Academy of
Sciences.
\end{theacknowledgments}



\end{document}